\documentstyle[epsf]{mn}

\title[Gravitational collisions in $N$-body codes]
{Gravitational collisions in cosmological $N$-body codes}
\author[I.~Suisalu and E.~Saar]
       {I.~Suisalu$^{1,2,3}$\thanks{E-mail: ivar@aai.ee}
	 and E.~Saar$^{1,3}$\thanks{E-mail: saar@aai.ee}\\
$^1$ Tartu Astrophysical Observatory, T\~oravere, EE-2444, Estonia\\
$^2$ Royal Observatory Edinburgh, Blackford Hill, Edinburgh EH9 3HJ, U.K.\\
$^3$ Theoretical Astrophysics Center, Blegsdamvej 17, Copenhagen, DK-2100,
Denmark
}

\begin{document}
\maketitle

\begin{abstract}
We study the accumulation of errors in cosmological $N$-body algorithms
that are caused by representing the continuous distribution of matter
by massive particles, comparing the P$^3$M and Adaptive Multigrid codes.
We use for this a new measure of two-body relaxation suitable
for nonstationary gravitating systems. This measure is based on accumulated
deflection angles of particle orbits and it does not saturate during evolution.
We have found that the role of gravitational collisions
in standard P$^3$M-models is rather high, and in order to avoid
collision effects we recommend to
use comoving softening parameters $\varepsilon>=1.0$ (in grid units).
\end{abstract}


\section{Introduction}

Cosmological simulations use $N$-body algorithms that
replace the initially continuous distribution of
(dark) matter by discrete and rather massive `particles'.
This helps to simulate intersecting streams of collisionless
matter that are common at the time of formation of
cosmological structure.
Interactions between these particles may, however, differ
from the dynamics of the continuous media they are meant
to approximate. The main source of discrepancy are
close encounters between the particles  (gravitational
collisions).

The popular belief has been that if the number of points
in the simulation is sufficiently large, the discreteness
effects (gravitational collisions or
close encounters between mass points)
do not play an important role. These effects have been studied
only for tree-codes in case of stationary gravitating
systems (Hernquist \& Barnes 1990, Huang, Dubinski \& Carlberg 1993).
We shall describe in this paper the role of gravitational
collisions in high-resolution cosmological codes.

Among the different N-body algorithms
used for cosmological simulations the
P$^3$M (Particle-Particle-Particle-Mesh)
algorithm has become the industry standard
for high-resolution simulations. It was
designed initially for plasma physics simulations and
ported to cosmology by Efstathiou \& Eastwood (1981).
It has been clearly documented in textbooks (Hockney \& Eastwood, 1988,
Couchman, 1995)
and the code is practically in public domain. This code was
the first candidate for our study.

The P$^3$M-code improves considerably the resolution of a
standard PM (Particle-Mesh) code by splitting forces in two parts --
the long-range force from the overall density distribution on a mesh
and the short-range force from nearest
neighbours, calculating the latter by summation over
pairs of particles. The algorithm was devised initially
to simulate collections of pointlike particles.
In cosmological simulations of formation of structure
the basic dark matter density is smooth and does not
lend itself easily to representation by discrete mass points.
This difficulty is usually alleviated by
introducing smoothing parameters in  pairwise forces,
effectively replacing the
mass points by extended spherical clouds.

In order to study dicreteness properties of the P$^3$M-code
one should have
smooth codes with spatial resolution comparable to the P$^3$M.
Such codes have appeared only recently, and most of them
are grid-based multi-resolution schemes (except that of
Pen (1994) that uses a global deformed grid). These codes
improve the spatial resolution in selected regions (mostly
in those of high density). We shall describe these codes
and their main differences below; for comparison we used
our adaptive multigrid code (Suisalu \& Saar 1995).

We ran also one simulation using the popular tree-code
(Barnes \& Hut 1986). Its deflection properties have been
studied before (Hernquist \& Barnes 1990, Huang et al. 1993),
but for a stationary case only.

The usual tool to measure pairwise collisions is the study
of energy diffusion (see the studies of three-code simulations
cited above).
This is proper for stationary gravitating
systems, where energy is otherwise conserved and heating comes
only from pairwise interactions; in cosmology the overall energy
changes with time and it is not the best measure for collisions.
We propose instead of it to study orbital deflection angles,
concentrating only on the change of direction of velocities
and not on the change of their absolute values.

\section{Simulations}

As the adaptive grid methods are not so well known
as the P$^3$M and one is forced to select between different
versions, we shall describe their differences below.

All these methods form subgrids of higher spatial resolution
in regions where density is higher than a prescribed limit.
They use for boundary conditions on a subgrid the
values of the potential (Dirichlet conditions) interpolated from
the coarser grid, but they differ when finding the solutions
for the potential.
The problem is if the coarse grid solution should depend
on the finer grid solution at the same point, or not. It certainly
should if we used exact direct solvers of linear
systems arising from discretized partial differential
equations, but the situation is not clear for approximate iterative
solvers. In cosmological simulations we do not have an exact
source term, either, as our density determination may depend
on a grid level.

In an ideal situation solutions of different resolution should
converge rapidly during the iteration process.
In practice convergence is rather slow, as
solutions on coarser grids are
considerably changed by temporary solutions on
finer subgrids (see, e.g. Brandt 1987). This issue has been
discussed by Anninos, Norman \& Clarke (1994)
who classify adaptive grid methods as having either
one or two-way interfaces between parent and subgrids.

As an example, in the multi-grid method developed by
Jessop, Duncan \& Smith (1994) only one-way
grid interfaces are used that implies that the local
fine grid solution can be different from the coarse grid solution.
They find the local solution by iteration
on the subgrid only. A recent adaptive code by
Splinter (1995) uses a similar methodology, although the algorithms
for finding the solution on subgrids differ.

In contrast, in
our AMG (Adaptive Multigrid)
code (Suisalu \& Saar 1995)
the solution for local subgrids is obtained simultaneously
with the global solution using the full multigrid method (Brandt 1977).
According to the  above classification AMG
uses two-way grid interfaces between finer and coarser grids,
as information
passes in both directions
during iterations. The reason why we have been
left alone in this class is that here it is harder to get
good convergence of the iterative solution process.
We believe that two-way interfaces, once they have been built,
are closer to the exact solution, and we shall use below
our AMG-code for comparison with the AP$^3$M-code.

At first we encountered difficulties in building our models,
as the natural boundary conditions for a multigrid code are the
Dirichlet conditions, and our code was initially tuned
for this case. In order to follow the evolution
of spatially periodic P$^3$M models we had to modify our code for
periodic boundary conditions.

If we use iterative methods to solve the Poisson equation on a grid,
we discover that in the case of periodic boundaries the linearized
system of algebraic equations that we have to solve becomes singular.
Singular systems are more difficult to solve, and the problem
becomes even harder if we consider interactions between the subgrids
and the global grid. In the case of full two-way interfaces
that we use, solutions on subgrids always induce changes in global
solutions that violate global boundary conditions, and this requires
usually additional iterations. In the periodic case the global
solution is very sensitive to violations of the zero total mass
condition that are generated by local subgrids, and convergence
becomes very slow.

The remedy we chose
was to use the Galerkin condition to modify the difference operators
on coarser grids. Namely, we calculate
the difference operator $L^H$ on the coarse grid as
\begin{equation}
L^H=I^H_h L^h I^h_H,
\end{equation}
where $L^h$ is the difference operator on the fine grid,
$I^H_h$ is the restriction operator from the fine grid to
the coarse grid and $I^h_H$ is the corresponding interpolation
operator. This technique helps to integrate singular
systems, and it is
described in detail by Hackbusch (1985).
We found it essential for speeding up the convergence of
the solution for the gravitational potential.

We chose to impose these
conditions only when finding the solution for the global grid
and did not modify the differential operator $L$ on subgrids.
In order to satisfy the boundary conditions themselves,
we copied appropriately the boundary
regions on global grids between iterations.

For P$^3$M simulations we used H.~Couchman's Adaptive P$^3$M
code AP$^3$M (Couchman 1991) that speeds up the normal
P$^3$M by generating subgrids in regions of high particle
density and finding there a smooth solution for the
potential. This works to decrease the volume of pairwise
force summation and considerably speeds up the algorithm.
As concerns the adaptive smooth solution,
the AP$^3$M-method belongs to the class of those with
an one-way interface.

\begin{table}
     \caption{Simulation parameters}
\begin{tabular}{@{}lrrrl}
\hline
Sim.&L&N&$\varepsilon$&$h_{min}$\\
\hline
P3M1$^\star$&$32^3$&$32^3$&0.2&1\\
P3M2&$32^3$&$32^3$&1.0&1\\
P3M3&$32^3$&$32^3$&2.0&1\\
P3M4&$32^3$&$64^3$&0.2&1\\
APM&$32^3$&$32^3$&0.2$^\dagger$&1/8\\
AMG1&$32^3$&$32^3$&&1/16\\
AMG2&$32^3$&$64^3$&&1/32\\
TREE&\multicolumn{2}{r}{16440}&0.2\\
\hline
\multicolumn{5}{l}{$^\star$ Three different spectra}\\
\multicolumn{5}{l}{$^\dagger$ Short-range accelerations excluded}\\
\hline
\end{tabular}
\label{models}
\end{table}

We have run and analyzed 6 $\rm{P^3M}$ simulations,
3 PM-type simulations and one tree-code simulation,
their parameters are summarized in Table~\ref{models}.
The first column
labels the simulations, $L$ is the base mesh size in cells
(the physical size is $80h^{-1}$ Mpc for all
simulations), $N$ is
the number of particles used,
$\varepsilon$ is the comoving softening parameter and
$\rm{h_{min}}$ the minimal
meshsize for adaptive models.

Couchman (1995) advises using softening radii
that are constant in physical coordinates that leads to
increasing discreteness during simulations. While this could
be appropriate to decribe the interaction of separate mass
concentrations that may form during the simulation,
it is certainly not the best description for dark
matter. We used comoving
softening to keep P$^3$M-simulations closer to
smooth grid simulations; e.g., Gelb \& Bertschinger (1994) have chosen
the same approach.

\begin{figure}
\epsfbox{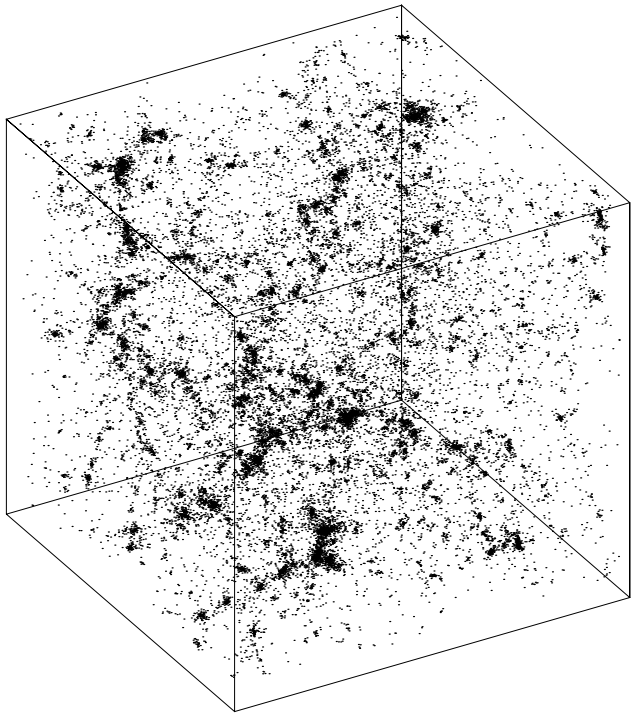}
\caption{
Particle distribution for the model P3M1 (P$3$M with
a small softening parameter $\varepsilon=0.2$) at
the present epoch $a=9$. The spectral index is $n=-1$ and the size of
the cube is 80$h^{-1}$ Mpc.}
\label{p3mpic}
\end{figure}

\begin{figure}
\epsfbox{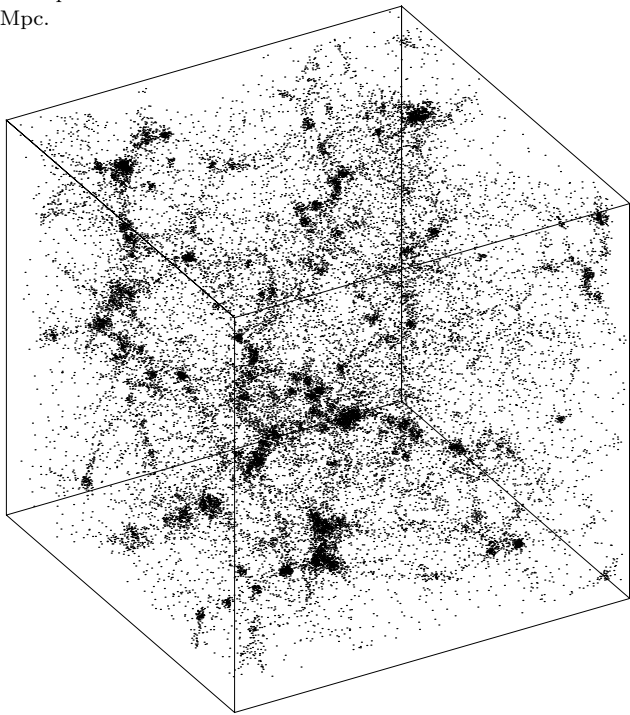}
\caption{
Particle distribution for the model AMG1 (AMG with
4 subgrid levels) at $a=9$. The initial conditions and the cube size are
the same as for the model P3M1 in Fig.~\protect\ref{p3mpic}.}
\label{amgpic}
\end{figure}

Altogether we used three different values of the
softening parameter $\varepsilon=0.2,1$ and 2 (our minimal
softening is that normally used in P$^3$M-simulations).
The maximum number of refinement levels for AP$^3$M was 4.
The model we call APM is similar to the
P3M1 with the only difference that we commented out in the
source code the lines that updated
velocities by  accelerations from short-range
forces -- this is essentially a PM-model with adaptive mesh refinement,
hidden inside the camp of AP$^3$M-models.

Two other PM
simulations labeled AMG1 and AMG2 were run using our
Adaptive Multigrid code modified for periodic boundary
conditions as described above. This code gives a spatial resolution
for forces similar to that of P$^3$M codes without the need to
consider pairwise forces.
The last two codes differ by the
number of points used and also by the number of subgrid levels
allowed, 4 for AMG1 and 5 for AMG2.

The tree-code we used was the
so-called `Barnes' export' code with quadrupole corrections. While
the computational volume for all other models was a periodic cube
with the size of $80h^{-1}$ Mpc, the volume for the tree-code was
a sphere (with the same diameter). We used an opening angle
$\theta=0.8$ and physical coordinates with the timestep of
0.001 Gy.

We used the same initial conditions for all our models,
trying to eliminate all possible sources of differences.
The initial conditions were generated using the test
power law spectrum with the spectral index $n=-1$
from the Couchman's Adaptive P$^3$M
distribution (see Couchman 1995 for a detailed description). For the
P3M1-model we generated also two other realizations with the initial
spectra as power laws with indices $n=-2$ and $n=0$. When we compare
different models from the Table~\ref{models}, we use always those
with the $n=-1$ spectrum.

As we simulated the simple $\Omega=1$ cosmology, we could
choose our starting time at will (we use the scalefactor
$a$ as our time variable). We started at $a=1$ and
followed the evolution until $a=9$ when $\sigma_8\approx1$
for all models, thus we brought our models up to the
present time. In order to study the growth of orbital
deflections further, we followed the evolution of selected
models for longer periods, up to $a=30$.

\begin{figure}
\epsfbox{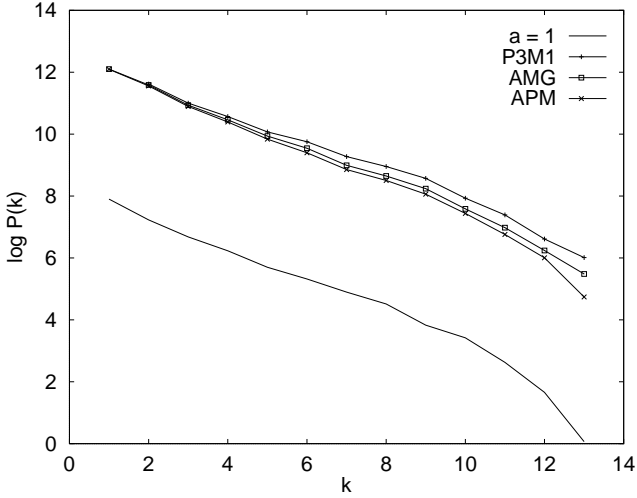}
\caption{
The evolution of the initial density power spectrum $P(k)$
(the curve marked $a=1$) in
different models. The spectra for the three models shown at the final
 moment $a=9$
are labeled according to Table~\protect\ref{models}. The wavenumber $k$
is given in the comoving grid units.}
\label{spectrum}
\end{figure}

We did not use the usual energy condition to check our time steps.
Instead of this we use in our AMG code
the Courant condition:
\begin{equation}
v_{max}\Delta a <  0.2h_{min},
\end{equation}
(we change the time step when necessary, using an
asymmetric leapfrog integration). As the AP$^3$M-code uses a constant
timestep, we ran first our AMG1-model, found the minimum
timestep used ($\Delta a = 0.0625$) and used it for all
simulations (we even ran AMG1 once more). This quarantees the
use of the same algorithm for integration in time and eliminates one
possible source of difference between models.
The value of $\Delta a$ we chose corresponds to
$\Delta p = 0.01$ for the time variable $p = (3/2)a$
used in AP$^3$M for $\Omega=1$. This time step
is much smaller than usual, but it is necessary
to follow accurately particle trajectories.

The typical density distribution for our models is shown
in Figs.~\ref{p3mpic} and \ref{amgpic}.
The first one shows the particle distribution for
the model P3M1 at $a=9$ and the second one a similar one for the
model AMG1. The density distribution in the former seems
to be more developed and has more distinct substructure, while
AMG has retained more of linear structure
elements and is less clumpy.
We may suspect that the difference is due to pairwise collisions but we
cannot say this on the basis of
comparing density distributions only.
The overall impression of both models
is rather similar, of course -- they start from identical
initial conditions.

The evolution of the spectrum for different models was also rather similar
(see Fig.~\ref{spectrum}). The difference is in the growth of small-scale
modes that is the highest for the P$^3$M-models with a small softening
(P3M1), the lowest for the APM model (this is essentially a pure PM-model)
and intermediate for the smooth but high-resolution AMG1 run. The difference
between the last two models can be explained by different depths
of adaptive subgrid levels used (3 for APM and 4 for AMG1) and by the fact
that the effective smoothing in P$^3$M is somewhat higher than that
in  AMG (about 3.5 versus 2 cell sizes). This makes the resolution achieved by
AMG1 about 4 times higher than that in APM.

\section{Deflection angles}

It is well known that during the linear stage of growth of
density perturbations particle velocities retain their
initial directions; the well-known Zeldovich approximation
says
\begin{equation}
\bmath{x}=\bmath{q}+b(t)\bmath{v}(\bmath{q}),
\end{equation}
where $\bmath{x}$ is the comoving coordinate of a particle labeled
by its initial coordinates $\bmath{q}$ and $b(t)$ is the density growth
rate from the linear theory, a function of time only. This shows clearly
that while the velocity amplitude may change in time, its direction
remains fixed. In fact, the components of the velocity direction are
adiabatic invariants of motion in the initial stage, while
particle energies change with time even without any
interaction. This suggests that the
change of the velocity direction is a better measure
of gravitational interaction than the usually used energy
diffusion.

Nonlinear effects -- growth of structure and pairwise collisions --
both contribute to the change of the direction of velocity, and if we
define the deflection angle ${\mathrm d}\phi$ for a particle by
\begin{equation}
{\mathrm d}\phi=|\bmath{v}\cdot\bmath{dv}|/v^2,
\end{equation}
then the total angle $\phi$ accumulated during the run of the simulation
reflects all gravitational collisions suffered by the particle.
The quantity $\phi$ does not describe the real angle between the initial
and final direction of the velocity of a particle, as it omits
the other degree of freedom $\theta$ necessary to describe a direction in
a 3-D space.
However, in case of a small total deflection the two angles are
connected by
\begin{equation} \label{standish}
\langle\sin^2\Phi\rangle=2/3[1-\exp(-3S/2)]
\end{equation}
(Standish \& Aksnes 1969), where the accumulated square
deflection for a trajectory $S$ is
\begin{equation} \label{Sdef}
S=\sum_a({\mathrm d}\phi)^2
\end{equation}
($a$ is our time coordinate).
The mean in (\ref{standish}) concerns
all trajectories with the same ${\mathrm d}\phi$-sequence but with
different $\theta$. For small values of $S$
it is equal to the mean square of the final deflection
angle $\Phi$, but for a long history of collisions $\langle\sin^2\Phi\rangle$
will reach the value 2/3 that describes an isotropic distribution.
On the contrary, the accumulated square deflection $S$ does not saturate
and continues to describe the total effect of gravitational
interactions. In this respect it is also better than the
often-used measure of orbital divergence that saturates easily
for spatially limited systems.

In order to be able to catch large single deflections we computed the
deflection angles in the simulations as
\begin{equation}
{\mathrm d}\phi = \arccos(\bmath{v}_t \cdot \bmath{v}_{t+\delta t}/
		|\bmath{v}_t||\bmath{v}_{t+\delta t}|),
\end{equation}
this did not add significantly to the cost of the simulations.
We summed both the deflection angles and their squares
for every trajectory during a simulation.

After this work was finished, we learned about a similar measure proposed
recently by Bagla \&
Padmanabhan (1995) to describe nonlinearity in cosmological
structure formation. They describe deviations from the linear stage
by a measure
\begin{equation} \label{bagla}
D_{gu}=(\bmath{u}-\bmath{g})^2/u^2
\end{equation}
(as $\bmath{u}$ is the velocity with respect to
the dimensionless time $a$, its dimension is the same
as that of the acceleration $\bmath{g}$). Their measure is easier to
calculate than ours, but it has to be modified for other cosmologies
(the Zeldovich approximation gives
zero for this expression only for $\Omega=1$ universe), ours does
not depend on the background cosmology.
But we see clearly that there are two sources
for orbital deflections -- nonlinear evolution of particle
orbits in the smooth background field described by (\ref{bagla})
and the pairwise (gravitational) collisions we are looking for. Of course,
both these sources contribute to the energy diffusion, too.

While deflection angles reflect histories of individual
particles, the
simplest characteristic describing the change of all particle trajectories
in a simulation is
the mean of accumulated squares of deflection
angles for all trajectories $\langle S\rangle$.
If particle motions are quasi-linear, as in
the Zeldovich approximation,
particles follow their initial direction (only their velocity
may change in time) and
$\langle S\rangle=0$. Growth of the deflection angles describes nonlinear
interactions, either via the mean field or by pairwise
gravitational collisions.

\begin{figure}
\epsfbox{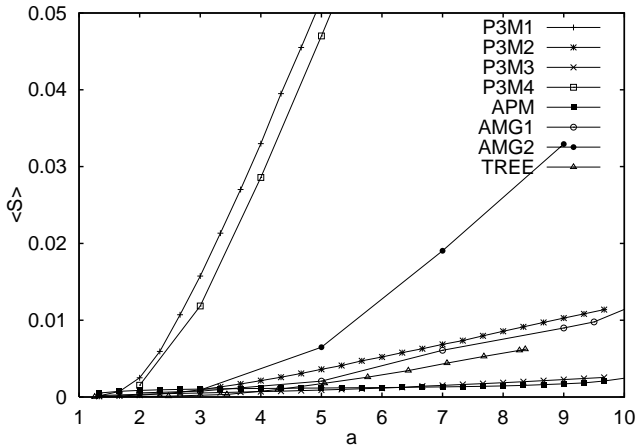}
\caption{
Growth of the mean accumulated square
deflection $\rangle S\langle$ with dimensionless time $a$ (the scale factor).
The models are labeled according to Table~\protect\ref{models}; the
highest growth is observed
in case of the P$^3$M-models with a normal smoothing parameter
$\varepsilon=0.2$.}
\label{S-all}
\end{figure}

In Fig.~\ref{S-all} we see the evolution of the deflection measure
$\langle S\rangle$ in time
for different models. First we see a striking difference
between the P$^3$M-models with different smoothing parameters;
for the normally accepted smoothing parameter ($\varepsilon=0.2$)
the accumulation of deflection angles is very rapid
compared to those for larger $\varepsilon$. Contrary to the
common belief, the rate of growth of deflection angles does not
depend much of the total number of particles $N$ -- compare the curves P3M1
and P3M4, the latter is for a model with 8 times more particles than the P3M1.
Our multigrid models
lie in the middle of the range, the model AMG1 practically coinciding with the
P3M2. This is understandable, as $\varepsilon=1$ corresponds to a
cell-sized smoothing; the model P3M3 $\varepsilon=2$ is evidently
`oversmoothed'.

\begin{figure}
\epsfbox{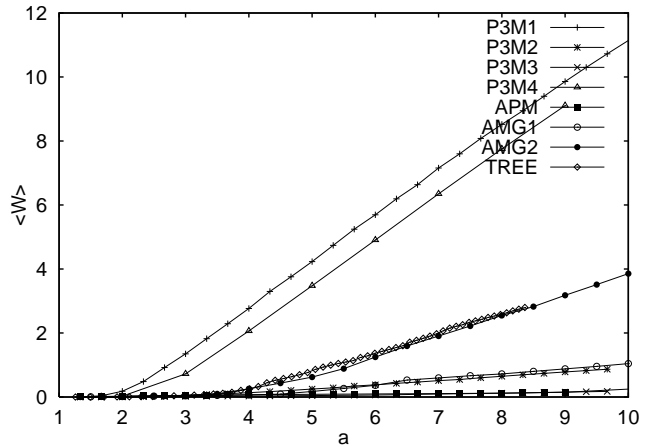}
\caption{
Growth of the mean squared angular velocity deflections
$\rangle W\langle$ with dimensionless time $a$. The models
are labeled according to Table~\protect\ref{models}; the highest growth
is observed
in case of the P$^3$M-models with a normal smoothing parameter
$\varepsilon=0.2$.}
\label{W-all}
\end{figure}

Surprisingly, the model AMG2 gives larger
deflections than the AMG1. This can hardly be caused by the larger number
of particles, a much more likely cause is that we
have allowed an extra level of refinement in this model and follow
particle trajectories better. An additional source of deflections
could be the force errors that each
refinement introduces near subgrid boundaries;
as we have shown earlier (Suisalu and Saar 1995), these errors are
usually small, less than one per cent, but in rare cases
they may reach a few per cent. Similar errors are present in the
AP$^3$M-code, where the usual requirement is to limit them by
6 per cent. We stress once more that this value describes very
rare errors. Anyway, such errors may get amplified
in a square-weighted characteristic as $\langle S\rangle$ is.
Another surprise is a relatively low deflection measure obtained
for our tree-code run.

As we mentioned above, the deflection may be caused both
by gravitational collisions and by interactions with the
mean gravitational field. It is well known than
in case of uncorrelated deflections particles follow a
random walk in the energy space (see, e.g., Huang et al.
1993). This would translate in our case to
\begin{equation}  \label{ranwalk}
{{\mathrm d}S\over{\mathrm d}a}=n(\Delta\phi)^2
\end{equation}
that would give a $\langle S\rangle\sim t$ dependence ($n$ is here the
mean frequency of collisions and $(\Delta\phi)^2$ the characteristic
value of a single squared deflection). This is close to
the $a$-dependence seen in Fig.~\ref{S-all}.

However, in our nonstationary simulations
we have to consider also the possible role of the mean-field
effects.
We can roughly model strong mean-field deflections,
supposing that all particles rotate in circular orbits with
angular velocities $\omega_i$. In this case the computed
deflection measure would grow as
\begin{equation} \label{rotation}
\langle S(a)\rangle={1\over N}\sum_i^N\sum_a(\omega_i\Delta a)^2=
	{a\Delta a\over N}\sum_i\omega_i^2.
\end{equation}
As we see, $\langle S\rangle$ caused by a strong mean-field
interaction is also proportional to time $a$, as the expected
effect of collisions was.

In order to differentiate between the signatures of mean-field
deflections and those caused by pairwise collisions,
we introduced a new (`velocity') deflection measure $W$
for the growth of the squared deflections of the angular velocity
of a particle:
\begin{equation} \label{Wdef}
W(a) = \int_{a_0}^a \left({{\mathrm d}\phi\over{\mathrm d}a} -
		\omega\right)^2 {\mathrm d}a,
\end{equation}
where $\omega$ is the
time-averaged rate of
change of the deflection angle $\omega$ for a particle
(the average angular velocity
for planar orbits) during the whole simulation (from $a_0$ to $a$):
\begin{equation} \label{omdef}
\omega=1/(a-a_0)\int_{a_0}^a {{\mathrm d}\phi\over{\mathrm d}a}{\mathrm d}a
\end{equation}
The velocity deflection measure $W$ is, of course, zero
for linear motions without the change of direction, but
also for the case of strong interactions,
when particles are trapped into systems and
rotate with constant angular velocities $\omega_i$,
$\phi_i = \omega_i a$. In order to describe all particles
we use a mean value of this measure $\langle W\rangle$.
The growth of this measure in time
for our simulations is shown in Fig.~\ref{W-all}.

The difference of the mean velocity deflection measure
between various models (Fig.~\ref{W-all}) is
considerably smaller than it was for the mean squared deflections
(Fig.~\ref{S-all}). The P3M1-model leads the pack as before,
with its many-particle version close behind, and the tree-code model
has generated similar deflections to the high-resolution AMG1.
All other models are relatively quiet.

\begin{figure}
\epsfbox{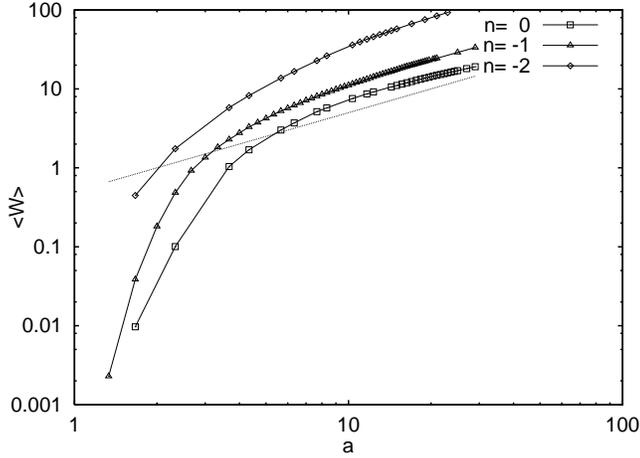}
\caption{
Growth of the velocity deflection measure $\langle W\rangle$
in the model P3M1 for different
initial spectra with exponents $n=0$, $n=-1$ and $n=-2$.
The dotted line shows a power law $\langle W\rangle\sim a$.}
\label{W-spectra}
\end{figure}

We studied the $\varepsilon=0.2$ case in more detail, trying to
better understand the source of deflections. We continued
the run more deep into the nonlinear times (Fig.~\ref{W-spectra}). We can see
that the growth, although rather rapid
at the start of the simulation (this rise could be probably
explained by non-self-consistent starting conditions),
levels off to a power
law with the exponent $\alpha\approx 1$.
As we have largely eliminated the mean-field effects, we
may hope that this exponent tells us that the observed
growth of $\langle W\rangle$ is caused by two-body effects.

In order to check if
this exponent depends on the large-scale environment
we repeated the simulations for the model P3M1 for two more
power spectra with the exponents $n=-2$ and $n=0$. As can be seen
from Fig.~\ref{W-spectra}, the deflections grow in a similar fashion, and the
exponents are practically the same -- the curves differ by a
multiplicative factor only. Examination of density distributions
confirms  that structure is much stronger in the $n=-2$ case, the
$n=0$ spectrum giving rise to a large number of small clumps and a
rather diffusive sea of particles in between. This will naturally
lead to a larger collision rate for the former model.

\begin{figure}
\epsfbox{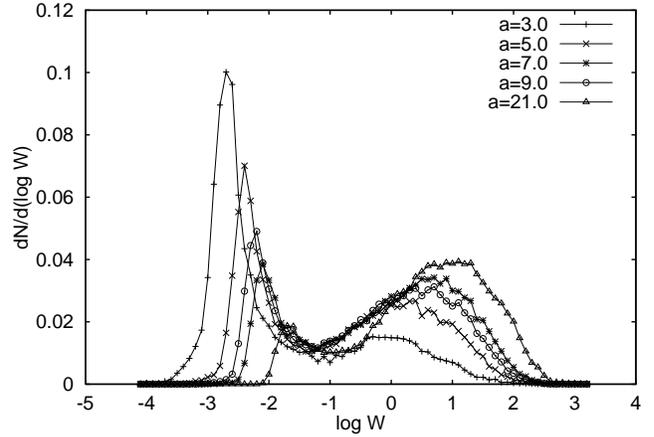}
\caption{
Distribution of the velocity deflection measure $W$ for all
particles in the P$^3$M-model P3M1 for different moments $a$.
The signature of high-deflection gravitational collisions is
clearly seen as the right maximum. Observe
how the distribution that is dominated by small deflections at
the start of the simulation shifts gradually over to relatively
large deflections.}
\label{Wdist-ppm}
\end{figure}

It is clear that different particles follow extremely different orbits,
and in order to understand their evolution it is better
to 'differentiate' the average quantitities shown in
Figs.~\ref{W-all}-\ref{W-spectra} and to
study the distributions of the velocity deflection measure for our
collection of trajectories.

Fig.~\ref{Wdist-ppm} demonstrates the evolution of
this distribution in time for the collisional model P3M1.
The main feature of this distribution is the presence of two strong maxima at
all times. We may suppose that the left maximum at smaller deflections
is describing mean-field effects and
the right maximum -- gravitational collisions. During the evolution
the number of particles that participate in collisions grows
steadily. We see also the gradual shift of the distributions
towards larger deflections -- the accumulated velocity deflections
grow. The width of the right maximum that can be thought of as describing
the number of strong collisions (about 3 per unit logarithmic
interval) grows also.

\begin{figure}
\epsfbox{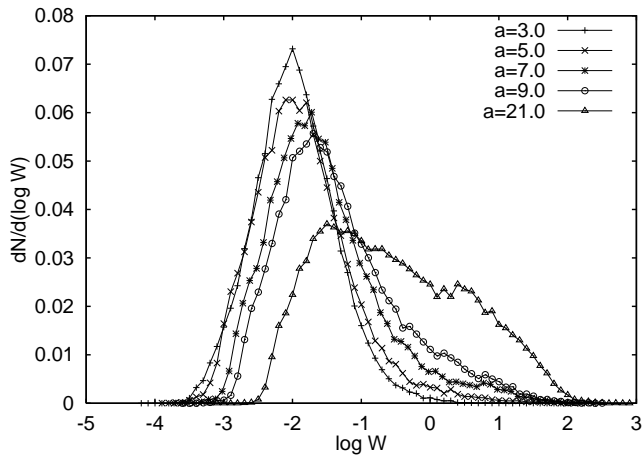}
\caption{
Distribution of the velocity deflection measure $W$ for the
multigrid model AMG1 for different moments. The deflection
measure grows gradually, but small deflections dominate at
all times.}
\label{Wdist-amg}
\end{figure}

Similar distributions for the multigrid model AMG1 are shown in
Fig.~\ref{Wdist-amg}.
They are distinctly different from the P$^3$M model shown before,
having only one small-deflection maximum. During
evolution the distribution spreads and only the last
distribution that corresponds to highly nonlinear stages of
evolution of structure (more than two present lifetimes of the Universe
into the future) shows presence of an appreciable
high-deflection tail. This tail is caused by the growth of small-scale
substructure by far beyond the present epoch.

\begin{figure}
\epsfbox{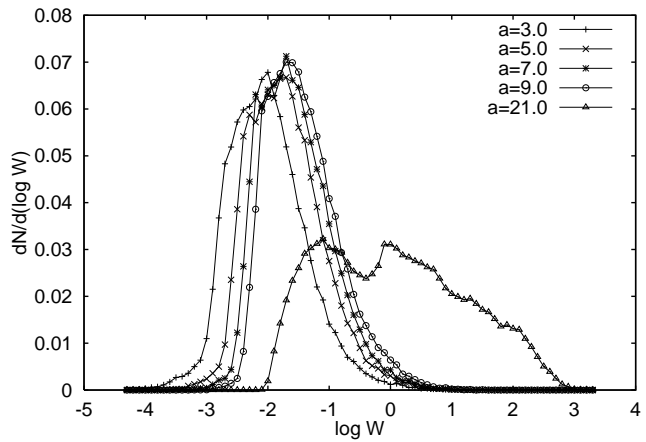}
\caption{
Distribution of the velocity deflection measure $W$ for the APM-model
(without local forces) at different times. The distribution is initially
smoother than for the multigrid models in Fig.~\protect\ref{Wdist-amg},
and the final highly nonlinear stage develops similar fluctuations.}
\label{Wdist-apm}
\end{figure}

Even more smooth distributions can be found when running the
APM model, where the local pairwise forces were ignored. The
distributions (Fig.~\ref{Wdist-apm}) show only the signature of
mean-field forces and
do not evolve practically at all. This shows, first, a total absence
of gravitational collisions typical for its parent P$^3$M-model, but
also less substructure than in the AMG1-model
(the spatial resolution of the present model
was about 4 times lower than that of the AMG1). An exception is the
distribution for large times that shows features similar to the AMG1,
even with slightly larger amplitudes.

\begin{figure}
\epsfbox{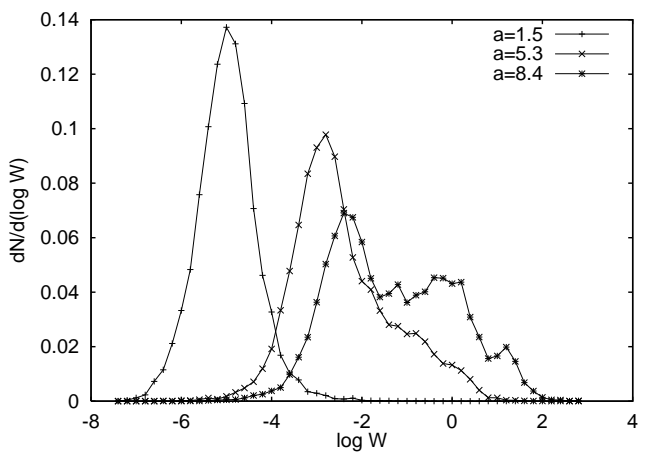}
\caption{
Distribution of the velocity deflection measure $W$ for the tree-code
simulation at different times. The distribution is rather wide,
but has only one maximum at all times..}
\label{Wdist-tree}
\end{figure}

The tree-code shows features in between of the P3M1-model and
the smooth models -- the distribution of the velocity deflection
measure $W$ (Fig.~\ref{Wdist-tree}) is rather wide, comparable
to that of the model P3M1, but it has only one maximum at all times.
As the transformation between the physical time coordinate $t$ used here
and the `scale factor time' $a$ used for other models is nonlinear,
it is difficult to compare these distributions directly with
the $W$-distributions for other models. Even the strong-field
approximations (\ref{rotation}) do not agree with each other,
and the measure $W$ here implies a non-constant mean angular
velocity in $a$.

\begin{figure}
\epsfbox{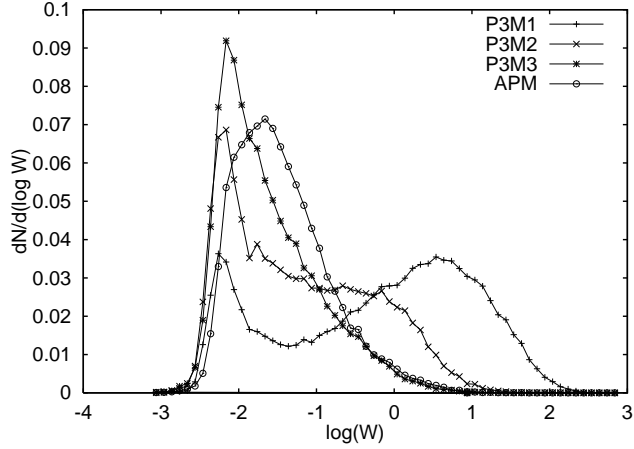}
\caption{
Distribution of the velocity deflection measure $W$ at the final
stage of simulations for P$^3$M-models with different softening
parameters. The role of collisions (the right maximum) decreases
rapidly with a growing $\varepsilon$.}
\label{Wdist-eps}
\end{figure}

As we saw above, the overall growth of the deflection measure
was similar for our multigrid models and for the P$^3$M-models
with rather large softening parameters (Fig.~\ref{S-all}).
The comparison of the distributions of the velocity deflection measure
$W$ at the end of the simulations, $a=9$ for the
P$^3$M-models with different softening is
shown in Fig.~\ref{Wdist-eps}. Only the standard model P3M1
($\varepsilon=0.2$) shows a strong collision signature, for the
smooth model P3M2 ($\varepsilon=1$) the right maximum is
already very weak and the `oversmoothed' P3M3 ($\varepsilon=2$)
shows only mean-field deflections. It is even smoother than
the APM-model, where the local pairwise interactions were
ignored. We see also that the models with a large softening
develop much smaller accumulated deflections.

\begin{figure}
\epsfbox{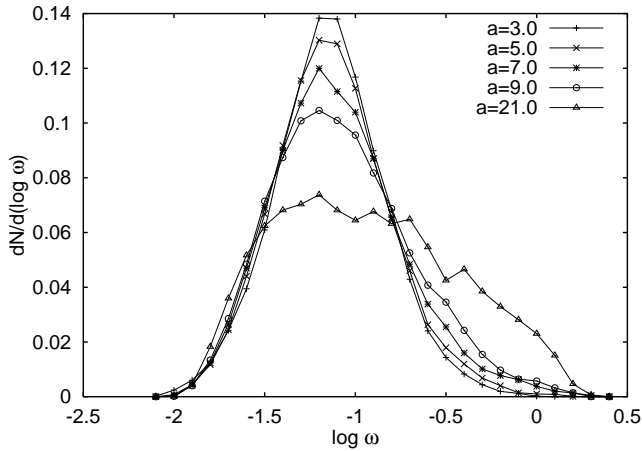}
\caption{
Distribution of the mean angular velocity $\omega$ for
different moments (the model AMG1). The initially concentrated distribution
spreads in time both due to collisions and mean-field effects.}
\label{omega}
\end{figure}

The deflection characteristics we have studied so far have
all been accumulated from the start of the simulation.
As the deflection angle is positively defined, the change of
its mean value, describing smooth-field effects, will
contribute in some extent to the velocity deflection parameter $W$.
We show a typical distribution of $\omega$ in Fig.~\ref{omega} for the
model AMG1. Its time dependence is rather slow, but $\omega$ tends
to grow, and this may influence the accumulated velocity
deflections.

In order to eliminate this effect, we performed another step
of `differentiation', separating the overall evolution into
a number of time intervals $a\in[a_i,a_{i+1}]$
and calculating the rate of growth
of the velocity deflection measure (deflection rate) $U$ by
\begin{equation}
U=W(a_i,a_{i+1})/(a_{i+1}-a_i),
\end{equation}
where $W(a_i,a_{i+1})$ is the same expression as in
(\ref{Wdef},\ref{omdef}) but we have changed the integration limits
from $(a_0,a)$ to $(a_i,a_{i+1})$. It is important to keep
in mind that this operation does not substract distributions
at different times, we differentiate along the accumulated
angular velocity deflections.

\begin{figure}
\epsfbox{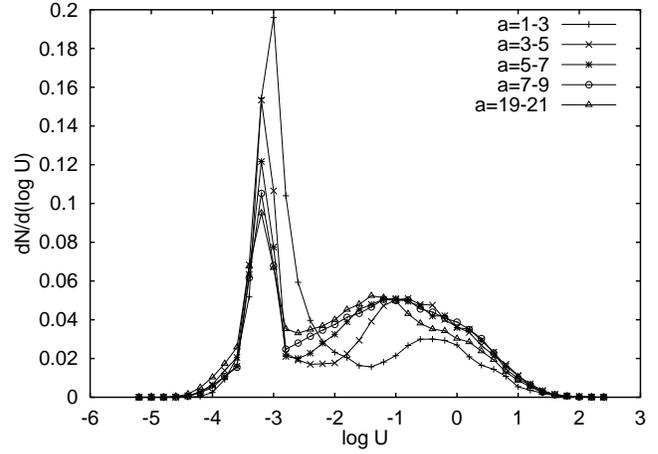}
\caption{
Distributions of the deflection rate
$U$ for the P$^3$M-model P3M1 for different time intervals.
The left maximum describes the mean-field effects while the right
maximum describing collisions grows steadily.}
\label{Udist-ppm}
\end{figure}

In Fig.~\ref{Udist-ppm} we see again the results for the P3M1-simulation.
The picture is essentially the same as in Figure~\ref{Wdist-ppm}. Only
the distributions for different times are more similar, and the roles of
small deflections for a subinterval (mean-field effects) and those of large
deflections
are more clearly separated. The small-scale peaks do not
move, indicating that the mean-field effects stay the same
during the evolution.
We built a similar graph for the model P3M4 that differs from
the present model only by a larger number of
particles. As suggested by a similar behaviour of the
deflection measure $S$ (see Fig.~\ref{S-all}) the distributions are
practically the same.

\begin{figure}
\epsfbox{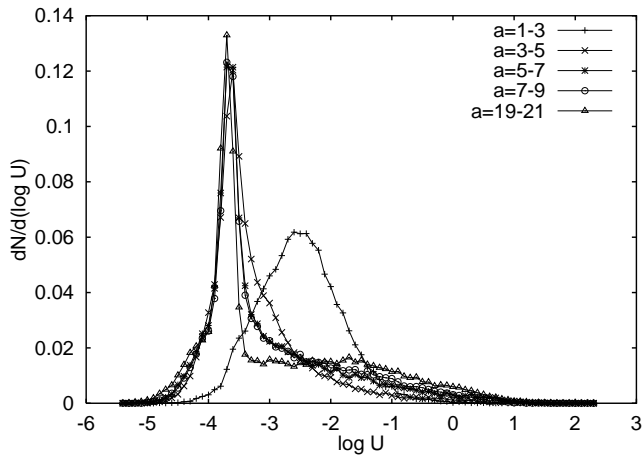}
\caption{
Distributions of the deflection rate
$U$ for the multigrid model AMG1. After the initial fast
evolution towards a quasiequilibrium qrowth, most
particles keep moving smoothly and only a small percentage of
them suffers deflections that place them in the
high-deflection tail of the distribution.}
\label{Udist-amg}
\end{figure}

The distributions of the deflection rate
in the adaptive multigrid code are shown in Fig.~\ref{Udist-amg}.
The figure reveals a noticable difference between the $U$-distribution
from the first and the successive time intervals. A probable reason
for this could be a non-self-consistent initial state that
gives rise to initial transients (such transients were
observed also by Hernquist \& Barnes, 1990, in their analysis of tree-code
models) . This could be caused by
the presence of softening in the force law as supposed
by Hernquist \& Barnes.  Models
evolve rapidly away from this stage, but they acquire in the process a
high-deflection tail of the distribution of the deflection
measures. The later
evolution is much more quiet, with most of the deflections
coming from small-amplitude mean-field accelerations.
At later times yet when higher resolution grids are used
we see also a growth of the high-deflection tail.
This could be partly a mean-field effect and could be partly
caused by force errors.
This tail is, however, much lower than that for the P3M1-model and there is
no sign of any maximum. This figure also shows that the strong difference
between the $W$-distribution for nonlinear stages (see Fig.~\ref{Wdist-amg})
is due to a large difference between the moments they have been constructed,
the deflection rate being practically the same at $a=9$ and at $a=20$.

\begin{figure}
\epsfbox{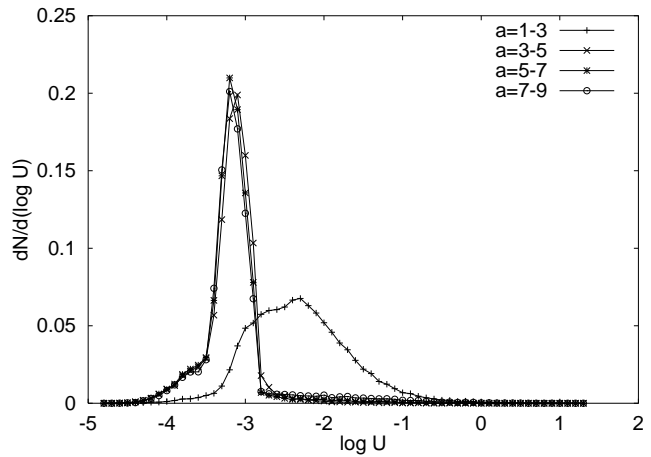}
\caption{
Distributions of the deflection rate
$U$ for the APM-model. Here one observes only
the initial fast relaxation, but there are practically
no collision events.}
\label{Udist-apm}
\end{figure}

It is instructive to compare Fig.~\ref{Udist-ppm}
with Fig.~\ref{Udist-apm}, where the $U$-distribution
from the APM simulation is shown.
This shows the difference between trajectories of particles
which undergo short-range accelerations and which do not. The APM
distribution is in fact very close to that for AMG, but
it almost does not have the long high-deflection tail.
There are at least two reasons for this -- firstly, the
adaptive grids in the APM did not reach as deep as they did in
the AMG-models, and secondly, the subgrid force calculation procedure
used in the AP$^3$M-approach might give cleaner forces.

Extra gravitational collisions inherent to a $N$-body code can
influence directly the conclusions that we make
comparing our models to observations. As they  generate
additional velocity changes, their presence is reflected most clearly in
the velocity distributions. These can be compared easily
with observations and they influence the properties of the
gravitationally bound systems that are born during the
simulations. As an example, we analyzed the
distributions of relative velocities for
particle pairs for different codes.

\begin{figure}
\epsfbox{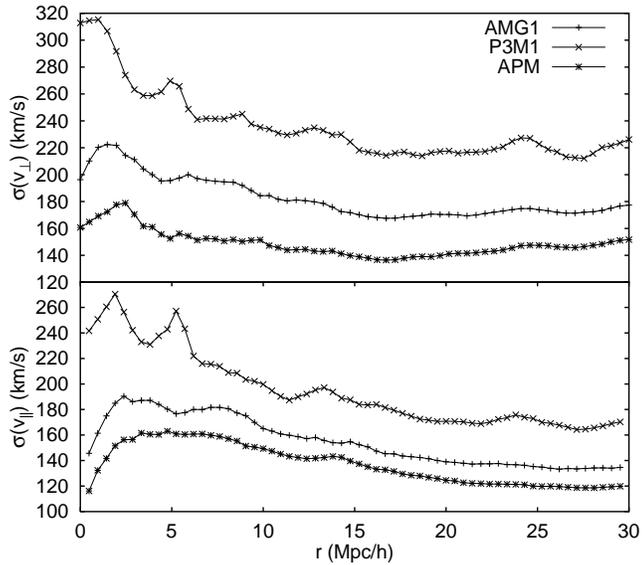}
\caption{
Dependence of the pairwise longitudinal (``radial'')
velocity dispersion $\sigma(v_\parallel)$ (panel a) and the
transversal
velocity dispersion (``rotation velocities'')
$\sigma(v_\perp)$
on the width of particle pairs $r$
for different models. We see that the dispersion obtained from
the P$^3$M-models is much larger than that from the smooth
models, even at relatively large distances.}
\label{veldisp}
\end{figure}

The relative velocity vector between a pair of particles can be decomposed
into the
line-of-sight (`radial') component and into the component perpendicular
to this direction (`tangential') (see Gelb 1992).
The radial component $v_\parallel$ can be defined as
\begin{equation}
v_\parallel = {(\bmath{v}_1 - \bmath{v}_2)\cdot(\bmath{x}_1 - \bmath{x}_2)
		\over |\bmath{x}_1 - \bmath{x}_2| }
\end{equation}
and the perpendicular velocity component $v_\perp$ as
\begin{equation}
v_\perp = ((\bmath{v}_1 - \bmath{v}_2)^2 - v_\parallel^2)^{1/2}.
\end{equation}

The dispersions of these velocities are shown in Fig.~\ref{veldisp}
for the final
moment $a=9$, corresponding roughly to the present epoch, for three
different models,
P3M1, AMG1 and APM. This
figure shows that the P$^3$M-code gives velocity
dispersions that are about 50\% larger than we get using smooth-field codes
to model the same patch of the Universe. The results we have seen above
let us suspect that this difference may be mainly due to two-body effects
in the P$^3$M-code. And as we already mentioned, this
could affect the conclusions one makes about the rotation velocities
and masses of the systems that form and about the ease they form with.

\section{Conclusions}

We have compared a range of P$^3$M-models with different softening
parameters and spectra and smooth-density
models of comparable spatial resolution (APM, multigrid).
We have seen that smooth-field PM-codes
are considerably less collisional than the P$^3$M-codes.
We have found that even choosing comoving softening parameters instead
of physical ones there is a considerable
amount of gravitational collisions in standard P$^3$M-models. If their
influence is crucial
for the problem the simulation is run for (e.g. the study of
velocity dispersions or of the formation of bound systems), we would
recommend to
use comoving softening parameters $\varepsilon>=1.0$.

We have also proposed a new approach to study gravitational collisions
 in nonstationary systems
that are common in cosmological simulations. We use measures
based on the accumulated deflection angles of particle orbits and
on accumulated angular velocity deflections. These measures are similar
to the commonly used energy diffusion and orbit divergence measures,
but they do not saturate during evolution.

\section*{Acknowledgements}

Our special thanks go to Dr. H.~Couchman who sent us his AP$^3$M code
that was one of the two workhorses for this study.
We thank Dr. J.~Peacock for suggesting the subinterval analysis for
the deflection velocity distributions that made the picture much clearer.
We are much indebted to Dr. B.~Jones for the discussion of the paper.
IS thanks the Edinburgh Observatory and the
Edinburgh Parallel Computing Centre, where
most of the simulations for this work were done, and both of us thank
the new Theoretical Astrophysics Center in Copenhagen, where the article
was finished, for supporting our stay there.
IS was supported in Edinburgh
by the EC TRACS scheme and both authors were supported by the
Estonian Science Foundation grant 338.

\end{document}